\documentclass[letterpaper,12pt]{article}
\usepackage{amsmath}
\usepackage{graphicx}
\usepackage{epstopdf}
\usepackage{url}
\usepackage{multirow}
\usepackage{authblk}
\usepackage{color}  
\usepackage{enumitem}

\usepackage{amssymb,amsfonts}
\usepackage{epsfig}
\usepackage{latexsym}
\usepackage{float}

\usepackage{hyperref}
\usepackage{xcolor}
\usepackage{lineno}

\usepackage[margin=0.65in]{geometry}

\usepackage{lineno}

\def \ba{\begin{eqnarray}}\def\ea{\end{eqnarray}}
\def\bc{\begin{center}}\def\ec{\end{center}}
\def\nn{\nonumber\\}

\title{\Large\bf Photoproduction of pion pairs  at high energy and small angles}

\author[1]{S.~Gevorkyan}
\author[2,3]{I.~Larin}
\author[2]{R.~Miskimen}
\author[4]{E.~Smith}

\affil[1]{Joint Institute for Nuclear Research, Dubna 141980, Russia}
\affil[2]{University of Massachusetts, Amherst, MA 01003, USA}
\affil[3]{Alikhanov Institute for Theoretical and Experimental Physics NRC Kurchatov Institute, Moscow 117218, Russia}
\affil[4]{Thomas Jefferson National Accelerator Facility, Newport News, VA 23606, USA}

\date{\today}

\begin{document}
\maketitle

\begin{abstract}
We explore the photoproduction mechanisms for charged and neutral pion pairs off a heavy nucleus at threshold.
We calculate the production of charged pairs in the Coulomb field of the nucleus in the Born approximation
using explicit expressions for the differential cross sections and their connection with the total cross section  $\sigma(\gamma\gamma\to\pi^+\pi^-)$.
The production of $\sigma$ mesons at threshold and their subsequent decay into pions is an important mechanism
and is investigated quantitatively in a framework that can be used for both charged and neutral pions.
$\sigma$ mesons can be produced via electromagnetic (photon exchange) or strong ($\omega$ exchange) interactions.
For neutral pions, another important production mechanism consists of the two-step process where charged pions are produced
first and then charge-exchange into two neutral pions. Differential and total cross sections are computed for each production mechanism.
\end{abstract}

\section{Introduction}
\begin{figure}[btp]
\centering
\includegraphics[width=0.25\linewidth,angle=0]{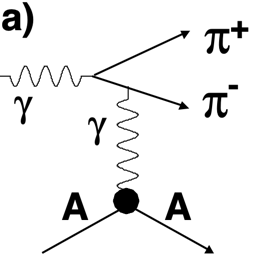}\hspace{3cm}
\includegraphics[width=0.25\linewidth,angle=0]{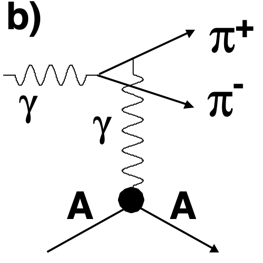}
\caption{Diagrams for the Primakoff production of charged pion pairs.}\label{fig1}
\end{figure}

\begin{figure}[btp]
\centering
\includegraphics[width=0.25\linewidth,angle=0]{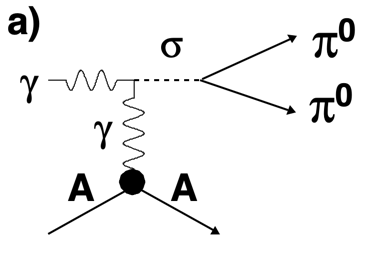}\hspace{1cm}
\includegraphics[width=0.25\linewidth,angle=0]{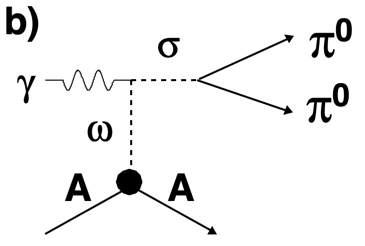}\hspace{1cm}
\includegraphics[width=0.25\linewidth,angle=0]{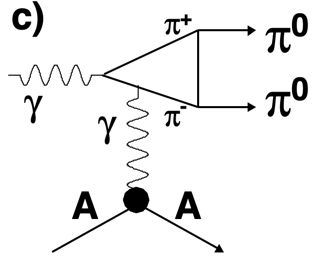}
\caption{Diagrams for the production of neutral pion pairs through 
a) Primakoff $\sigma$-meson production and decay, b) strong $\sigma$-meson production and decay, and c) production via charge exchange.}\label{fig2}
\end{figure}

The  photoproduction of  charged pion pairs on the  proton  $\gamma\,p\to\pi^+$$\pi^-\,p$ has
been  widely investigated beginning from the  seminal experiments at SLAC~\cite{Ballam:1971yd}.
At high energies $E_{\gamma}\gtrsim$ 3$\,$GeV, the pairs are primarily produced diffractively
(Pomeron exchange) in the $P$-wave (orbital momenta $L$$=$$1$) producing $\rho\,(770)$ vector mesons
and the $f_2(1270)$ tensor mesons ($D$-wave), which are easily observable in the $\pi^+\pi^-$
invariant mass spectrum. In addition, the charged pion pairs may be produced by photon
or vector ($\omega$, $\rho$) meson exchanges with the target nucleus (Fig.$\,$\ref{fig1}).
Such pairs are produced in the $S$-wave ($L=0$) resulting in scalar mesons:
$f_0$(500) and $f_0$(980). The threshold region is dominated by the $f_0$(500) meson, which we will
refer to as the $\sigma$-meson in this paper.
The production of neutral pion pairs is a result of scalar meson production
decaying to $\pi^0\pi^0$ (Fig.$\,$\ref{fig2}a,\,b) or the production of
charged meson pairs with subsequent charge exchange
($\pi^+\pi^-\to\pi^0\pi^0$, Fig.$\,$\ref{fig2}c).

Understanding double pion photoproduction at threshold is crucial for experiments
whose goal is to measure the pion polarizability using the Primakoff production
of two pions \cite{CPPexp,NPPexp}. In addition, it can also be an important background
to single meson Primakoff production experiments of $\eta, \eta\prime$ with the goal of
determining their lifetimes with high precision \cite{PrimexEta}.
Therefore, this paper presents the formalism and calculations for electromagnetic
and strong production of two pions in the field of a nucleus.
We start by estimating
the contribution to the photoproduction cross section of charged pions in the Coulomb field
and then present the relevant expressions that describe the production of neutral pion pairs.
   
\section{Photoproduction of charged pion pairs in the Coulomb field}
The reaction of interest can be expressed as
\ba
\gamma(k)+ A(p)\to \pi^+(q_+)+\pi^-(q_-)+ A(p'),
\ea
where the four-vectors of each particle are given in parentheses.
The differential  cross section for the photoproducion of charged
pion pairs in the Coulomb field  of nucleus (Fig.$\,$\ref{fig1})
can be obtained using the techniques from Ref.\,\cite{Gevorkyan_2012} for $M\gg 2\,m$,
where $m$ and $M$ are the single pion and pion pair invariant masses:
\ba
d\sigma=\frac{8\alpha^3Z^2}{(2\pi)^2}\frac{d^2q}{Q^4}\bigg (\vec q_-({\vec q_+}^2+m^2)+\vec q_+({\vec q_-}^2+m^2)\bigg )^2
\frac{\beta_-(1-\beta_-)d\beta_-d^2q_-}{(\vec q_-^2+m^2)^2(\vec q_+^2+m^2)^2}|F_{em}(q,q_L)|^2.\label{eqn2}
\ea
Here and below we are using the following notation:

$k = (E,\vec{\bf{k}})$ are the incident photon 4-momentum, energy and 3-momentum;

$k' = (E',\vec{\bf{k'}})$ are the pion pair 4-momentum, energy and 3-momentum;

$\Omega,\theta$ are the solid and polar angles of the produced pion pair;

$\vec{\bf{q}}_{+,-}$ the transverse part of the positive/negative pion 3-momenta
relative to the incident photon direction;

$\vec{\bf{q}} =\vec{\bf{q}}_-+\vec{\bf{q}}_+$ the transverse part of the transfer momentum;

$q_L=$ \scalebox{1.41}{$\frac{M^2}{2k}$} is the longitudinal transfer momentum;

$F_{em}(q,q_L)$ is the nucleus electromagnetic form factor;

$\beta_-$ is the fraction of photon energy taken by the $\pi^-$;

$Z$ is the charge of the nucleous;

\begin{figure}[H]
\centering
\includegraphics[width=0.9\linewidth,angle=0]{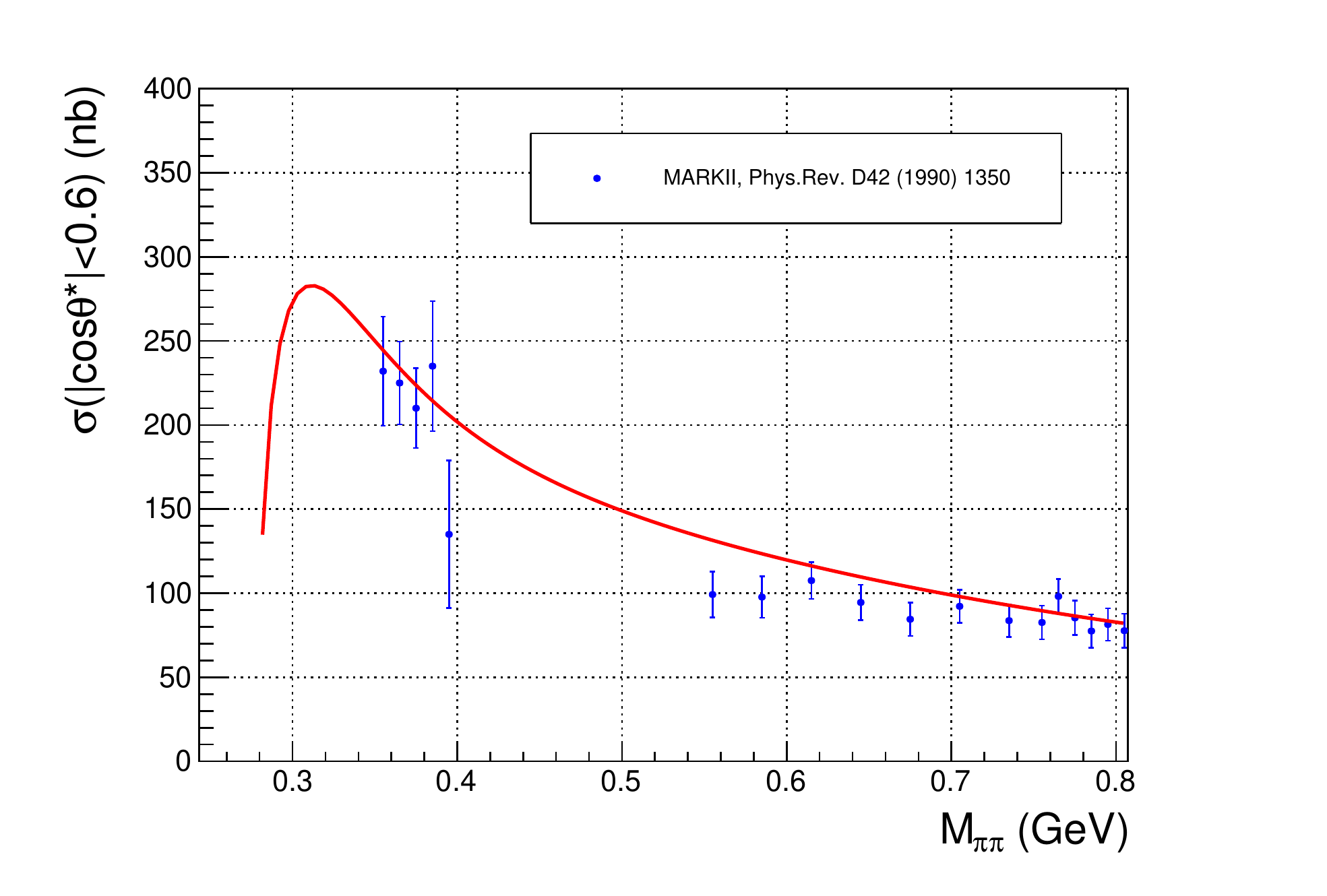}
\caption{The total cross section of the process $\sigma(\gamma\gamma\to\pi^+\pi^-)$
(equation \ref{form5}), scaled by factor of 0.6. The points with errors are the experimental data
for the solid angle range $|\cos(\theta^*)|<0.6$ from the Ref.~\cite{Boyer:1990vu}~(Table III).}\label{figf5}
\end{figure}

$Q^2=\vec q^2+ q_L^2$;\\
and $t=(k-k')^2$ is the Mandelstam variable.\\
In  the Primakoff region, where the transverse nuclear recoil can be safely neglected,
$\vec q^2$ is small and one can approximate ${\vec{q_-}}^2\approx {\vec{q_+}}^2$,
the expression simplifies to
\ba
d\sigma=\frac{8\alpha^3Z^2}{(2\pi)^2}\frac{\vec q^2d^2q}{Q^4}\frac{\beta_-(1-\beta_-)d\beta_-d^2q_-}
{({\vec q_-}^2+m^2)^2}|F_{em}|^2.
\ea
Recalling  that the invariant mass of the pair $ M^2=\frac{{\vec q_-}^2+m^2}{\beta_-(1-\beta_-)}$
and replacing $d^2q$ with $\pi dt$
we obtain:
\ba
\frac{d\sigma}{dt}=2\alpha^3Z^2\frac{\vec q^2}{Q^4}\frac{dM^2}{M^4}|F_{em}|^2.
\label{eqn4}
\ea
This expression was derived under the assumption of large pion pair mass M, where the asymptotic value for 
$\sigma(\gamma\gamma\to\pi^+\pi^-) = 2\pi\alpha^2/M^2$. A more accurate calculation is obtained by
using the full expression for the cross section as
a function of pair invariant mass~\cite{BUDNEV1975181}: 
\ba
\sigma(\gamma\gamma)=\frac{2\pi\alpha^2}{M^2}\left[\sqrt{1-\frac{4m^2}{M^2}}
(1+\frac{4m^2}{M^2})-\frac{4m^2}{M^2}(2-\frac{4m^2}{M^2})
\ln\left(\frac{M}{2m}+\sqrt{\frac{M^2}{4m^2}-1}\right)\right].     \label{form5}
\ea
We note that the cross section $\sigma(\gamma\gamma\to\pi^+\pi^-)$
is affected by higher order corrections due to the polarizability
(see for example Ref.~\cite{CPPexp}). This calculation is compared to data in  Fig.$\,$\ref{figf5} where the calculation
is scaled by a factor of 0.6.\footnote{This naive scaling factor accounts for the first-order correction due to the limited kinematic coverage of 
the experimental data.}
%
Equation~\ref{eqn4} can be rewritten as:
\ba
\frac{d\sigma}{dt}=\frac{2\alpha Z^2}{\pi}\frac{\vec q^2}{Q^4}\frac{dM}{M}\sigma(\gamma\gamma)|F_{em}|^2. \label{eqn7}
\ea
Equations~\ref{eqn2}--\ref{eqn4} and \ref{eqn7}) were obtained for $M\gg 2m$,
but the differential cross section of the pair production can be expressed via the
$\sigma(\gamma\gamma\to\pi\pi)$ as in Eq.\,\ref{eqn7} for any $M$ values~\cite{BUDNEV1975181, Halprin:1966zz}.
Eq.~\ref{eqn7} coincides with the equation (5.49) in Ref.~\cite{BUDNEV1975181}.
Finally, substituting $dt$ with \scalebox{1.41}{$\frac{|\vec{\bf{k}}||\vec{\bf{k'}}|}{\pi}$}$d\Omega$,
we obtain:
\ba
\frac{d\sigma}{d\Omega dM}=\frac{2\alpha Z^2}{\pi^2}\frac{|\vec{\bf{k}}|^2|\vec{\bf{k'}}|^2
 sin^2{\theta}}{Q^4}\frac{1}{M}\sigma(\gamma\gamma)|F_{em}|^2.
\ea
Thus, the photoproduction of charged pion pairs in the Primakoff region
has the same angular dependence as in the case of single $\pi^0$ meson
photoproduction~\cite{Miskimen:2011zz}, while dependence on invariant mass
of the pair $M$ is determined by the cross section of the process $\gamma\gamma\to\pi^+\pi^-$.

\section{Neutral pion pair photoproduction in the Coulomb field}

 Neutral pion pairs can be produced via a two step process:
 the photoproduction of scalar resonance $f$, followed by its decay
 to two pions (Fig.$\,$~\ref{fig2}a). The cross section of this process
\ba
\gamma(k)+ A(p)\to f+ A\to \pi^0+\pi^0+ A(p'),\label{eqn9}
\ea
can be written in the factorized form as
\ba
\frac{d\sigma}{dtdM^2}=\sum_f\frac{1}{\pi}\frac{d\sigma_f(t)}{dt}
\frac{m_f\Gamma(f\to\pi^0\pi^0)}{(M^2-m_f^2)^2+m_f^2\Gamma_{tot}^2}.\label{eqn10}
\ea
Here \scalebox{1.41}{$\frac{d\sigma_f(t)}{dt}$} is the differential  cross section of the scalar meson $f$
off the nucleus $\gamma+A\to f+A$; $\,\Gamma(f\to\pi^0\pi^0)$ and $\Gamma_{tot}$
are its partial and total decay widths, respectively; $M$ is two pion invariant mass,
and $m_f$ is $f$-meson mass.
We adopt cross section normalization as
\scalebox{1.41}{$\int{\frac{1}{\pi} \frac{d\sigma}{dtdM^2}}$}
$dM^2=B_r$ \scalebox{1.41}{$\frac{d\sigma}{dt}$},
where $B_r$ is the branching fraction of the $f$ meson decay to $\pi^0\pi^0$.
The use of the Breit-Wigner formula in Eq.$\,$\ref{eqn10} is a poor approximation for the
$f_0(500)$ meson because of its large width ($\Gamma_{tot}\sim\,$0.500$\,$GeV).
However, this representation is illustrative and can be easily substituted
with a more realistic description without affecting the basic results of the present work.
We note that the $f_0(500)$, which is also referred to as the $\sigma$ meson,
dominates the cross section at small invariant masses~\cite{CrystalBall:1990oiv}.

The differential cross section of $\sigma$ meson photoproduction in the nuclei Coulomb field is similar to $\pi^0$ photoprodction and  reads:
\ba
\frac{d\sigma_f}{dt}=\Gamma(f\to \gamma\gamma)\frac{8\pi\alpha Z^2\beta}{m_f^3}\frac{\vec q^2}{Q^4}|F_{em}|^2,\label{eqn15}
\ea
where $\beta$ is the $\sigma$ meson velocity~\cite{Gevorkyan_2012,Kaskulov:2011ab}.
Thus the photoproduction cross section for the $\sigma$ meson has the same angular dependence
as the well known~\cite{Miskimen:2011zz} photoproduction  cross section  of  the $\pi^0$ with replacement
of relevant masses and decay widths.          
Substituting this expression in Eq.\,\ref{eqn10} we get  the differential cross section of two $\pi^0$ photoproduction
in the Coulomb field of  nucleus:
\ba
\frac{d\sigma}{dtdM^2}=\frac{8\alpha Z^2\beta}{m_{\sigma}^2} 
\frac{\vec q^2}{Q^4}\frac{\Gamma(\sigma\to \gamma\gamma)\Gamma(\sigma\to\pi^0\pi^0)}
{(M^2-m_{\sigma}^2)^2+m_{\sigma}^2\Gamma_{\sigma tot}^2}|F_{em}|^2.\label{eqn16}
\ea
This expression allows one to calculate the neutral pion pair photoproduction in the Coulomb field of the nucleus
for any target, invariant mass of the pair, and transfer momentum values.
The meson  radiative  width can be written as
\ba
\Gamma(f\to\gamma\gamma)=(\frac{m_f}{4\pi})^3(\frac{\alpha}{F_f})^2N_c^2<e_f^2>^2,\label{eqn13}
\ea
where $N_c$ the number of colors, $ <e_f^2>$ the average of relevant quark charges and $F_{\sigma}=F_{\pi}=92.4 MeV$ is the charge meson decay constant.
For the $\pi^0$ meson  $N_c<e_{\pi}^2>=3\left((2/3)^2-(1/3)^2\right)=1$, whereas for the $\sigma$ meson
$N_c<e_{\sigma}^2>=3\left((2/3)^2+(1/3)^2\right)=\frac{5}{3}$.
The radiative decay width $\Gamma(\sigma\to\gamma\gamma)$ calculated by this expression yields $\Gamma(\sigma\to\gamma\gamma)=1.1$ keV.
However, for our calculations we have adopted the more reliable estimate of $\Gamma(\sigma\to\gamma\gamma)=2.05$~keV~\cite{Dai:2014lza}. For the $\sigma$ decay to $\pi^0\pi^0$ partial width, we use the value of $\Gamma(\sigma\to\pi^0\pi^0)=0.210\,$GeV~\cite{Agaev:2018sco}.

\section{Neutral pion pairs via the charge-exchange reaction}
As it was mentioned above, the $\pi^0\pi^0$ pair can be produced as a result of charge pion pair
photoproduction followed by charge exchange reaction $\pi^+\pi^-\to \pi^0\pi^0$.
The expression in Eq.\,\ref{eqn7} for the differential cross section of charge pion pair photoproduction
is also valid for the neutral pion pairs photoproduction if one replaces the $\sigma(\gamma\gamma)$
with the total cross section $\sigma(\gamma\gamma\to\pi^0\pi^0$) of the two step process
$\gamma\gamma\to\pi^+\pi^-\to\pi^0\pi^0$.
This cross section has been calculated in the framework of the Chiral theory\cite{Donoghue:1988eea} and gives:
\ba
\sigma(\gamma\gamma\to\pi^0\pi^0)&=&\frac{\alpha^2}{8\pi^2}\left(1-\frac{4m^2}{M^2}\right)
\left(1+\frac{m^2}{M^2}f(M^2)\right)\sigma(\pi^+\pi^-\to\pi^0\pi^0)\nn   
\sigma(\pi^+\pi^-\to\pi^0\pi^0)&=&\frac {1}{32\pi M^2 F_{\pi}^4}(M^2-m^2)^2,\nn
\mathrm{where:~} f(M^2)&=& 2\left[\ln^2(z_+/z_-)-\pi^2\right]+\frac{m^2}{M^2}\left[\ln^2(z_+/z_-)+\pi^2\right]^2,\nn
 z_\pm&=&\frac{1\pm(1-4m^2/M^2)^{1/2}}{2}.
\ea
Substituting this expression in Eq.\,\ref{eqn7}, one obtains the differential photoproduction cross section
of two neutral pions generated by the  charge-exchange reaction. We note that this contribution is about 2 orders of magnitude smaller
than the Coulomb production in the threshold region (see Fig.\,\ref{fig3cs}).

\section{Vector mesons exchanges in $\pi^0\pi^0$ photoproduction}
The photoproduction of neutral pion pairs in S-wave can also proceed through vector-meson exchange in addition
to photon exchange. However, we only consider $\omega$-meson exchange to the coherent photoproduction on nuclei
because the contribution from isospin one exchange ($\rho$ meson) is small and can be safely neglected.
To estimate the contribution of $\omega$ exchange, we could proceed as in the case
of $\pi^0$ meson photoproduction~\cite{PrimEx:2010fvg}, by substituting $\Gamma(\sigma\to \gamma\gamma)$
with $\Gamma(\sigma\to\omega\gamma)$ and
replacing the photon propagator with the $\omega$ propogator in Eq.\ref{eqn16}.
An alternative approach, which we implement here, is to use the simple parmetrization from Ref.\,\cite{Gevorkyan:2009ge},
which was used
previously to calculate the the strong amplitude for $\omega$ exchange to the production of $\pi^0$ photoproduction
on the nucleon
with $f_N(\gamma N\to\pi^0N)=L(\frac{E}{k_0})^{1.2}sin{\theta}$, where  $L=10\sqrt{\mu b}, k_0=1GeV$~\cite{Browman:1974cu}.
For the strong amplitude of $\sigma$ meson photoproduction, one can follow the same procedure
in Ref.\,\cite{Gevorkyan:2009ge} by
scaling the amplitude with the coefficient of 5/3 from Eq.\,\ref{eqn13}.
With this modification, the strong part of the differential cross section of $\sigma$ meson photoproduction
off a nucleus with atomic mass A in the Primakoff region reads as follows:
\ba
\frac{d\sigma_{str}}{dt}= A^2\frac{25\pi}{9E^2}f_N^2|F_{str}(q,q_L)|^2\label{eqn17}
\ea
The full differential cross section for neutral pion pair photoproduction
can be written similarly to the case of $\pi^0$ photoproduction in the Primakoff region as:
\ba
\frac{d\sigma}{dtdM^2}=\frac{1}{\pi}\left(\frac{d\sigma_{em}}{dt}+\frac{d\sigma_{str}}{dt}+
\sqrt{\frac{d\sigma_{em}}{dt}\frac{d\sigma_{str}}{dt}}\cos{\varphi}\right)
\frac{m_{\sigma}\Gamma(\sigma\to\pi^0\pi^0)}{(M^2-m_{\sigma}^2)^2+m_{\sigma}^2\Gamma_{\sigma tot}^2}\label{eqn18}
\ea
where {\large $\frac{d\sigma_{em}}{dt}$} and  {\large $\frac{d\sigma_{str}}{dt}$} are given by the expressions
in Eqs.\,\ref{eqn15} and \ref{eqn17} and
$\varphi$ is the relative phase between the Coulomb  and strong amplitudes.
The difference of the strong form factor $F_{str}(q,q_L)$ from the $\pi^0$ case~\cite{Gevorkyan:2009ge}
is minor. One has to replace the pion mass in the longitudinal momenta transfer $q_L$
with the $\sigma$ meson mass and also replace the total cross section
with nucleon $\sigma(\pi N)\to\sigma(\sigma N)$.

With these modifications, the electromagnetic and strong form
factors of the nucleus read as follows~\cite{Gevorkyan:2009ge}:
\ba
F_{em}(q,q_L)&=& 2\pi\frac{q^2+q_L^2}{q} \int_{0}^{\infty} J_1(q b)\,b^2\,db
 \int_{-\infty}^{+\infty}\frac{e^{i q_L z}}{(b^2+z^2)^{3/2}}\,dz\nn
&\times& \exp\left(-\frac{\sigma\prime A}{2}\int_{z}^{\infty}{\rho(b,z\prime)dz\prime}\right)
\int_{0}^{\sqrt{b^2+z^2}}{x^2\rho(x)dx} \label{eq:ffem}
\ea
\ba
F_{str}(q,q_L)=-\frac{2\pi}{q}\int_{0}^{\infty}J_1(q b)\,b\,db
 \int_{-\infty}^{+\infty}\frac{\partial\rho(b,z)}{\partial b}
  e^{i q_L z}\,dz\,\exp\left(-\frac{\sigma\prime A}{2}\int_{z}^{\infty}\rho(b,z\prime)dz\prime\right) \label{eq:ffstr}
\ea
Here $\rho(b,z)$ is the nucleus density;  $J_1(x)$ is the first order Bessel function;
$\sigma'=\sigma_{tot}(\sigma N) (1-i\frac{Re f(0)}{Im f(0)})$; 
$f(0)$ is the forward amplitude of $\sigma+N\to \sigma+N$; the invariant transfer momenta:
$t=-q^2-q_L^2 = -4\,|\vec{\bf{k}}|\, |\vec{\bf{k'}}|\, sin^2(\frac{\theta}{2})-(\frac{m_{\sigma}^2}{2E})^2$.
Substituting  expressions from Eqs.\,\ref{eqn15} and \ref{eqn17} for the electromagnetic and
strong cross sections into Eq.\,\ref{eqn18} one can calculate the momentum transfer and
invariant mass dependence of the differential cross section for double $\pi^0$ production off a nucleus.

\begin{figure}[H]
\centering
\includegraphics[width=0.5\linewidth,angle=0]{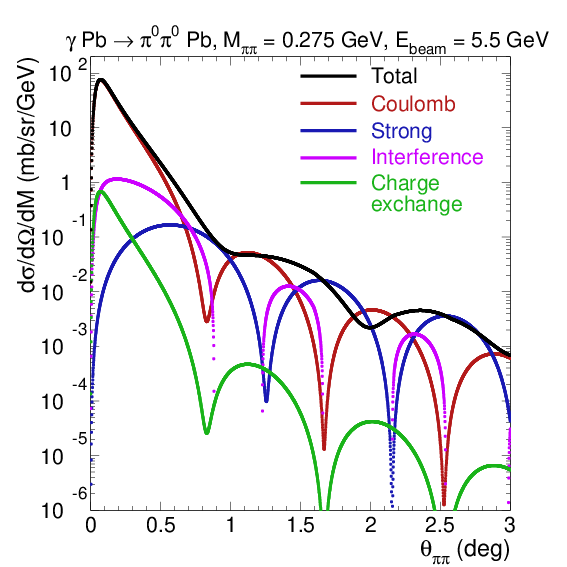}
\caption{Double differential cross section {\large $\frac{d^2\sigma}{d\Omega dM}$} of two
neutral pion photoproduction and its components
as a function of two pion production angle for lead-208 nucleus, 0.275\,GeV pion pair mass,
and 5.5$\,$GeV photon beam energy.
Red\,--\,Coulomb cross section for the amplitude shown in Fig.\,2a;
blue\,--\,strong cross section (Fig.\,2b); magenta\,--\,cross section from the interference of
Coulomb and strong amplitudes for 1\,radian interference angle;
green\,--\,cross section for the charge exchange amplitude shown in Fig.\,2c;
black\,--\,the total cross section, which is the sum of all contributions.}\label{fig3cs}
\end{figure}

\begin{figure}[H]
\centering
\includegraphics[width=0.5\linewidth,angle=0]{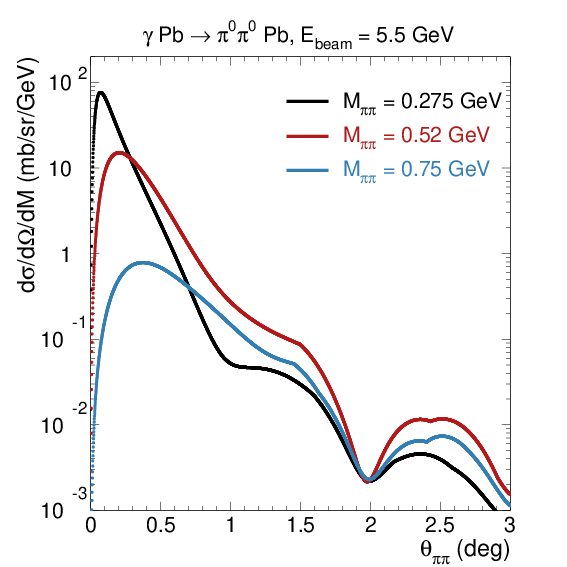}
\caption{Total double differential cross section 
{\large $\frac{d^2\sigma}{d\Omega dM}$} of neutral pair pion photoproduction
as a function of the pion pair production angle for lead-208 nucleus,
5.5$\,$GeV photon beam energy, and interference angle between Coulomb and strong mechanisms 1\,rad.
Black curve\,--\,0.275\,GeV pion pair mass, red\,--\,0.52\,GeV, blue\,--\,0.75\,GeV.}\label{fig3cs3m}
\end{figure}

\begin{figure}[H]
\centering
\includegraphics[width=0.5\linewidth,angle=0]{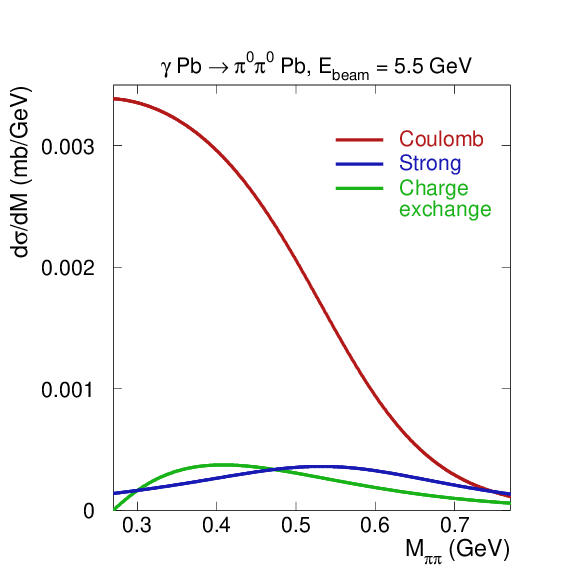}
\caption{Components of the differential cross section {\large $\frac{d\sigma}{dM}$}
of neutral pion pair photoproduction as a function of the two-pion invariant mass for lead-208 nucleus,
5.5$\,$GeV photon beam energy (integrated over production angle).
Red\,--\,Coulomb cross section for the amplitude shown in Fig.\,2a;
blue\,--\,strong cross section (Fig.\,2b);
green\,--\,cross section for the charge exchange amplitude shown in Fig.\,2c.}\label{dsdm1}
\end{figure}

Figures~\ref{fig3cs},~\ref{fig3cs3m},~and~\ref{dsdm1} show differential cross sections as a function of the pion pair
production angle and invariant mass obtained with the presented calculations for lead-208 nucleus and a
5.5$\,$GeV photon beam energy. For numerical calculations, we take the interference angle
between the Coulomb and strong mechanisms
to be $\phi=1$\,rad, which is close to the value measured for single $\pi^0$ production \cite{PrimEx-II:2020jwd}.

\section{Summary}
In this paper we presented explicit calculations for the most important production mechanism for pion pairs at forward angles and at threshold.
The production of $\pi^+\pi^-$ pairs was calculated in the Born approximation. The production of $\pi^0\pi^0$ pairs at threshold is dominated by 
the production and decay of the $\sigma$ meson, which has been calculated for electromagnetic  (photon exchange) and strong ($\omega$ exchange) mechanisms.
An additional production mechanism for neutral pion pairs proceeds via the two-step process where two charged pions are produced and
then rescatter into two neutral pions, which is found to be approximately 2 orders of magnitude smaller than the Coulomb contribution in the threshold region. 
Differential and total cross sections are computed for each production mechanism using realistic form factors in all cases.

Recently there have been several experimental programs aimed at measuring mesons at very small production angles in photoproduction 
in Hall D at Jefferson Lab. An ongoing experiment is measuring the Primakoff production of $\eta$ mesons with the goal of determining its lifetime to high precision \cite{PrimexEta}. Experiments to measure the Primakoff production of charged \cite{CPPexp} and neutral \cite{NPPexp} pion pairs 
are scheduled to run at Jefferson Lab in the near future. Therefore, quantitative estimates of the production of pion pairs at threshold and small angles 
will be important for the experimental program. In the first case these pairs represent background and in the second case they are the signal.


\section{Acknowledgements}
This material is based upon work supported by the U.S. Department of Energy, Office of Science,
Office of Nuclear Physics under contract DE-AC05-06OR23177, and UMass DOE contract DE-FG02-88ER40415.

\newpage
\bibliographystyle{unsrt}                                                                              
\bibliography{photoprod_pp}

\newpage
\appendix
\section{Numerical calculations and visualization}
\subsection{Primakoff cross section}
The Primakoff cross section proceeds via photon exchange for the process $\gamma(k)+ A(p)\to f+ A\to \pi^0+\pi^0+ A(p')$.
Let us calculate the two dimensional distribution of $\vec{q}^2, M^2$ using the expression
\ba
\frac{d\sigma}{d\Omega dM} = \frac{2 M |\vec{\bf{k}}||\vec{\bf{k'}}|}{\pi} \frac{d\sigma}{dtdM^2},
\ea
where $d\sigma/dtdM^2$ is given by Eq.\,\ref{eqn16}. We have used the following numerical values in the calculations:
$$\beta=1; m_{\sigma}=0.5GeV; Q^2=\vec{q}^2+q_L^2;
 \vec{q}^2=4\,|\vec{k}|^2\,|\vec{k'}|^2\,\sin^2{\theta/2}\approx (E\theta)^2; q_L=\frac{m_{\sigma}^2}{2k}$$
$$0.275GeV \leq M\leq 0.675GeV; 0\leq q^2\leq 0.01GeV$$
$$\Gamma_{\sigma tot}=0.5\,GeV;~~\Gamma(\sigma\to\gamma\gamma)=2.05\cdot10^{-6}\,GeV;
~~ \Gamma(\sigma\to\pi^0\pi^0)=0.21\,GeV$$.

\begin{figure}[btp]
\centering
\includegraphics[width=0.675\linewidth,angle=0]{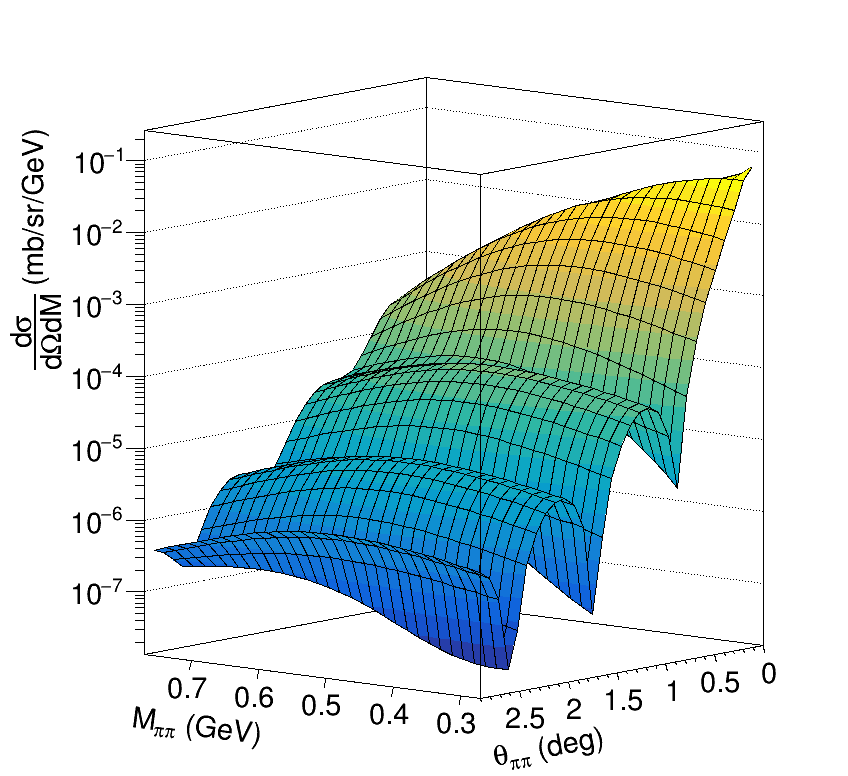}
\caption{Double differential cross section {\large $\frac{d^2\sigma}{d\Omega dM}$}
of the Coulomb neutral pion pair photoproduction
as a function of the pair production angle and invariant mass for lead-208 nucleus, and
5.5$\,$GeV photon beam energy.}\label{d2sdodmeps}
\end{figure}

The Coulomb double differential cross section $d^2\sigma/d\Omega dM$ is plotted in Fig.~\ref{d2sdodmeps}.
and the electromagnetic  form factor is given by Eq.\,\ref{eq:ffem}. The upper limit of integration 
goes to infinity since the Coulomb field covers all space.
For numerical calculations we note that outside of the nucleus the absorption
exponent is a constant. Thus we can take the exponent term outside the integral and this allows us to use 
the following analytical formulas to integrate the remaining function
\ba
\int_{0}^{\infty} \frac{cos(ax)}{(b^2+x^2)^{3/2}} dx = \frac{a}{b}K_1(ab);
\int_{0}^{\infty} \frac{sin(ax)}{(b^2+x^2)^{3/2}} dx = \frac{\pi}{2}\frac{a}{|b|}\{L_{-1}(ab)-I_1(|ab|)\},\nonumber
\ea
where $L_{-1}$ is the modified Struve function, and $K_1$ and $I_1$ -- the corresponding Bessel functions.
In this approach, we will need to subtract the difference
between this constant term and the actual value of the absorption exponent within the nucleus,
which has a limited size and thus convenient for the purpose of numerical calculations.
Using this method it is also easy to estimate the necessary region of integration for the variable $b$
to get precise form factor values. For large pion pair masses, we need to integrate $b$ up to about 50 -- 80\,fm
for convergence.
For the pion pair masses near threshold, we need to integrate to about 700 --1000\,fm to achieve a similar
level of convergence.\footnote{Electron screening may play
a role at the sub-percent level at such large distances from the nucleus.}

\begin{figure}[btp]
\centering
\includegraphics[width=0.425\linewidth,angle=0]{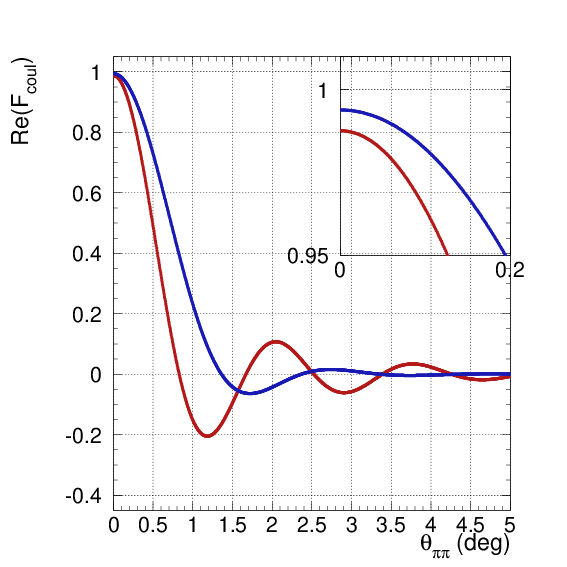}
\includegraphics[width=0.425\linewidth,angle=0]{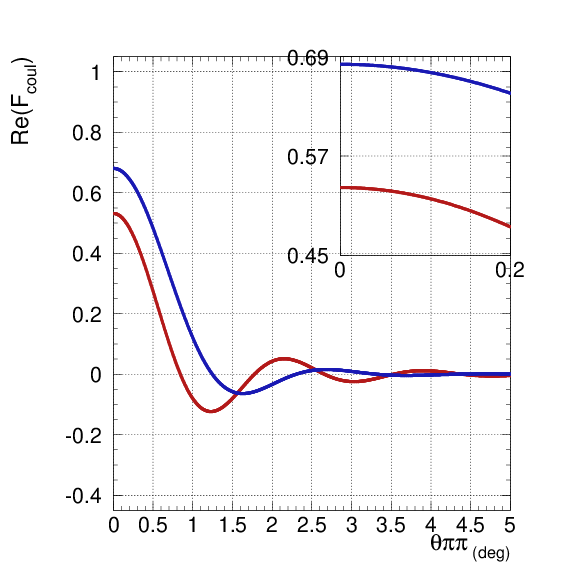}
\includegraphics[width=0.425\linewidth,angle=0]{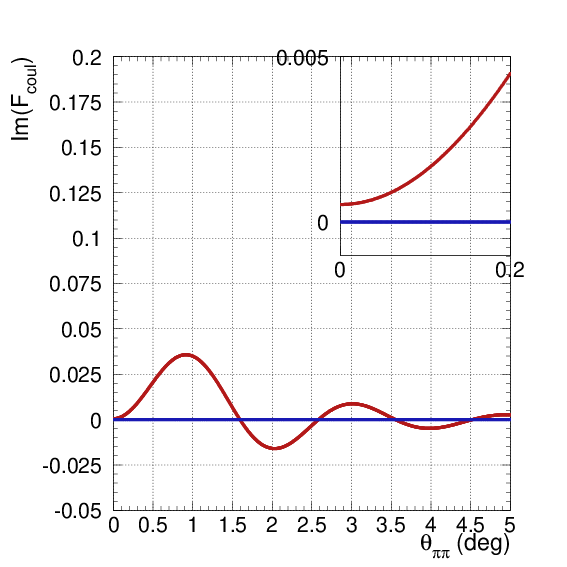}
\includegraphics[width=0.425\linewidth,angle=0]{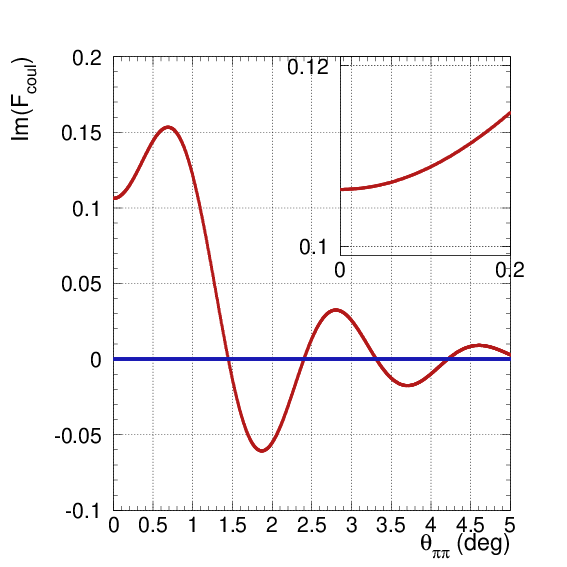}
\caption{Real (top plots) and imaginary (bottom plots) part
of the Coulomb form factor as a function of production angle
for the invariant mass values: 0.275\,GeV (left plots)
and 0.765\,GeV (right plots). Red curves -- full calculation,
blue curves -- no absorption.
The function behavior near the zero zoomed in the insertions.
\label{ffcoul}}
\end{figure}

The plots in Fig.\,\ref{ffcoul} show the real and imaginary parts of the Coulomb form factor as
a function of the production angle for $M_{\pi\pi}=$0.275\,GeV and
0.765\,GeV with and without including absorption. As seen from the plots, the presence of
the absorption not only decreases the value of the real part of the
form factor values but also enhances the oscillatory behavior.
We note that the interaction cross section in the absorption term is actually a function of the pion energy
and depends on the energy split between the pions in the pair. Including this
dependence properly could change our calculations by more than a percent.


\subsection{Electromagnetic and Strong cross sections} 
\begin{figure}[btp]
\centering
\includegraphics[width=0.675\linewidth,angle=0]{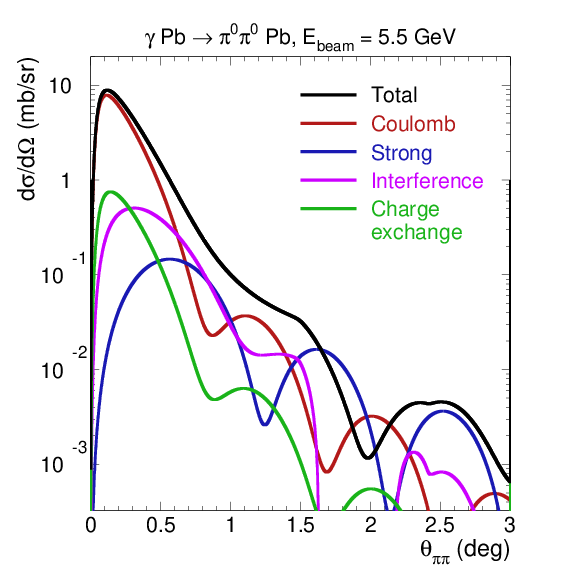}
\caption{Differential photoproduction cross section {\large $\frac{d\sigma}{d\Omega}$}
 of neutral pion pairs for each component
as a function of two pion production angle for a lead-208 nucleus,
5.5$\,$GeV photon beam energy, and integrated over the pion pair mass range between 0.27 and 0.77\,GeV.
Red\,--\,Coulomb part, blue\,--\,strong, magenta\,--\,their interference for 1 radian interference angle
(negative values are omitted on the log scale), green\,--\,charge exchange, black\,--\,total sum.
\label{dsdoeps}}
\end{figure}

\begin{figure}[btp]
\centering
\includegraphics[width=0.425\linewidth,angle=0]{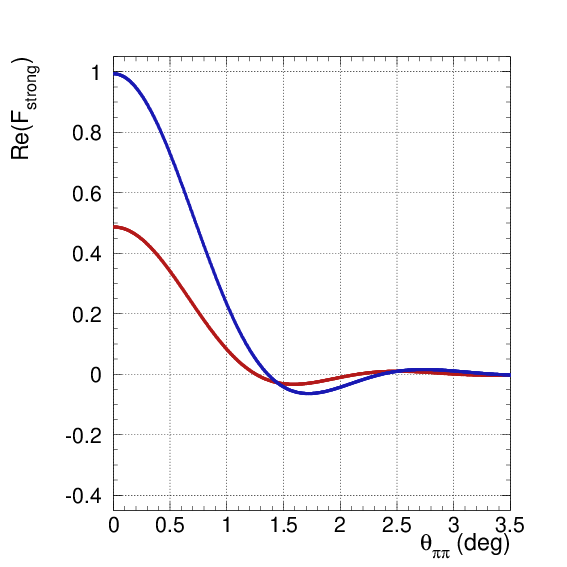}
\includegraphics[width=0.425\linewidth,angle=0]{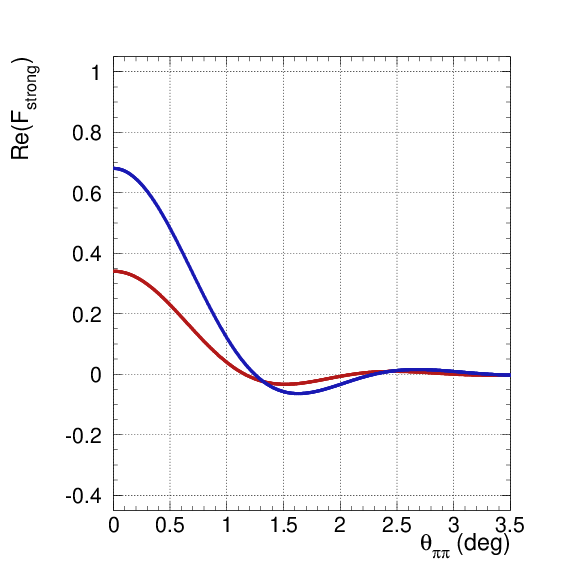}
\includegraphics[width=0.425\linewidth,angle=0]{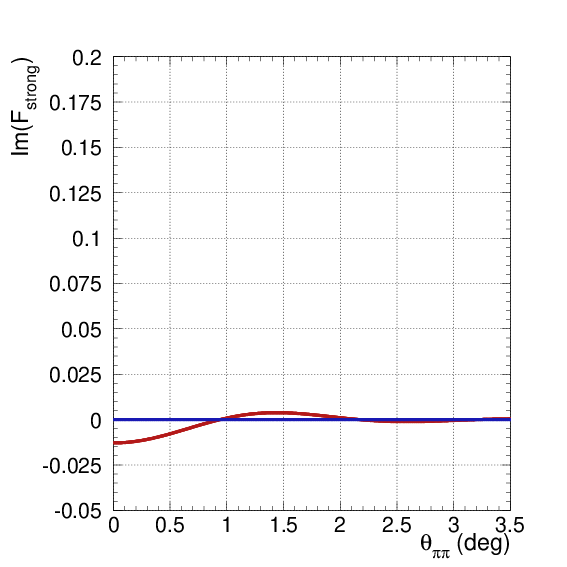}
\includegraphics[width=0.425\linewidth,angle=0]{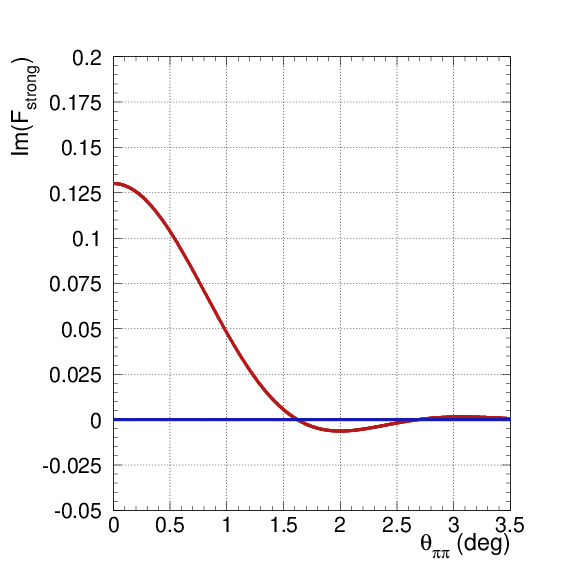}
\caption{Real (top plots) and imaginary (bottom plots) part
of the strong form factor as a function of production angle
for the invariant mass values: 0.275\,GeV (left plots)
and 0.765\,GeV (right plots). Red curves -- full calculation,
blue curves -- no absorption.
\label{ffstro}}
\end{figure}

The electromagnetic and strong cross sections integrated over invariant mass:
\ba
\frac{d\sigma}{d\Omega} = \frac{|\vec{\bf{k}}||\vec{\bf{k'}}|}{\pi} \frac{d\sigma}{dt};~~~
\frac{d\sigma}{dt}=\frac{d\sigma_{em}}{dt}+\frac{d\sigma_{str}}{dt}+
\sqrt{\frac{d\sigma_{em}}{dt}\frac{d\sigma_{str}}{dt}}\cos{\varphi}, 
\ea
where the expressions for the two components can be found in Eqs.\,\ref{eqn15} and \ref{eqn17}.
They are computed for $E=5.5\,GeV$, and the strong form factor is given by Eq.\,\ref{eq:ffstr}.
For the pion pair invariant mass integration range of 0.27--0.77\,GeV, the resulting cross section
{\large $\frac{d\sigma}{d\Omega} = \int_{0.27\,GeV}^{0.77\,GeV}\frac{d^2\sigma}{dM d\Omega}$}
shown in Fig.\,\ref{dsdoeps} for each term separately as well as for the total. 
The strong form factor as a function of production angle value is plotted in Fig.\,\ref{ffstro}.


\end{document}